\documentclass[conference]{IEEEtran}
\IEEEoverridecommandlockouts

\usepackage{cite}
\usepackage{algorithmic}
\usepackage{array}
\usepackage{url}
\usepackage{amsmath}
\usepackage{amsfonts}
\usepackage{amssymb}
\usepackage{tikz}
\usepackage[T1]{fontenc}
\usepackage{textcomp}

\usepackage{graphicx}
\usepackage{xcolor}
\def\BibTeX{{\rm B\kern-.05em{\sc i\kern-.025em b}\kern-.08em
    T\kern-.1667em\lower.7ex\hbox{E}\kern-.125emX}}

\begin{document}

\title{IoT DoS and DDoS Attack Detection using ResNet

\thanks{[© 2020 IEEE. Personal use of this material is permitted. Permission from IEEE must be obtained for all other uses, in any current or future media, including reprinting/republishing this material for advertising or promotional purposes, creating new collective works, for resale or redistribution to servers or lists, or reuse of any copyrighted component of this work in other works.]}

\author{\IEEEauthorblockN{Faisal Hussain,Syed Ghazanfar Abbas}
\IEEEauthorblockA{\textit{Al-Khawarizmi Institute of Computer} \\
Science (KICS) Lahore, Pakistan \\
faisal.hussain.engr@gmail.com, \\
ghazanfar.abbas@kics.edu.pk}
\and
\IEEEauthorblockN{ Muhammad Husnain,Ubaid U. Fayyaz}
\IEEEauthorblockA{\textit{Al-Khawarizmi Institute of Computer} \\
Science (KICS) Lahore, Pakistan \\
muhammad.husnain@kics.edu.pk, \\
ubaid@uet.edu.pk}
\and
\IEEEauthorblockN{Farrukh Shahzad,Ghalib A. Shah}
\IEEEauthorblockA{\textit{Al-Khawarizmi Institute of Computer} \\
Science (KICS) Lahore, Pakistan \\
farrukh.shahzad@kics.edu.pk, \\
ghalib@kics.edu.pk}
}
}

\maketitle
\begin{abstract}
The network attacks are increasing both in frequency and intensity with the rapid growth of internet of things (IoT) devices. Recently, denial of service (DoS) and distributed denial of service (DDoS) attacks are reported as the most frequent attacks in IoT networks. The traditional security solutions like firewalls, intrusion detection systems, etc., are unable to detect the complex DoS and DDoS attacks since most of them filter the normal and attack traffic based upon the static predefined rules. However, these solutions can become reliable and effective when integrated with artificial intelligence (AI) based techniques. During the last few years, deep learning models especially convolutional neural networks achieved high significance due to their outstanding performance in the image processing field. The potential of these convolutional neural network (CNN) models can be used to efficiently detect the complex DoS and DDoS by converting the network traffic dataset into images. Therefore, in this work, we proposed a methodology to convert the network traffic data into image form and trained a state-of-the-art CNN model, i.e., ResNet over the converted data. The proposed methodology accomplished 99.99\% accuracy for detecting the DoS and DDoS in case of binary classification. Furthermore, the proposed methodology achieved 87\% average precision for recognizing eleven types of DoS and DDoS attack patterns which is 9\% higher as compared to the state-of-the-art.

\end{abstract}


\section{Introduction}
Internet of Things (IoT) is the wireless interconnection of smart devices or things connected over the internet. In recent years, IoT has emerged as a promising technological solution for providing connectivity to myriads of heterogeneous devices across the globe. IoT can help us to access, control and manage these devices to get various functionalities in multiple application scenarios like smart home, smart healthcare, smart transportation, smart industry, etc. It can allow us to automate device control in order to facilitate the ease of device usage, to provide comfort and convenience to human being thus enhancing the overall quality of life.

\begin{figure}[t]
 \centering
  \centerline{\includegraphics[width=\linewidth, height=4.5cm]{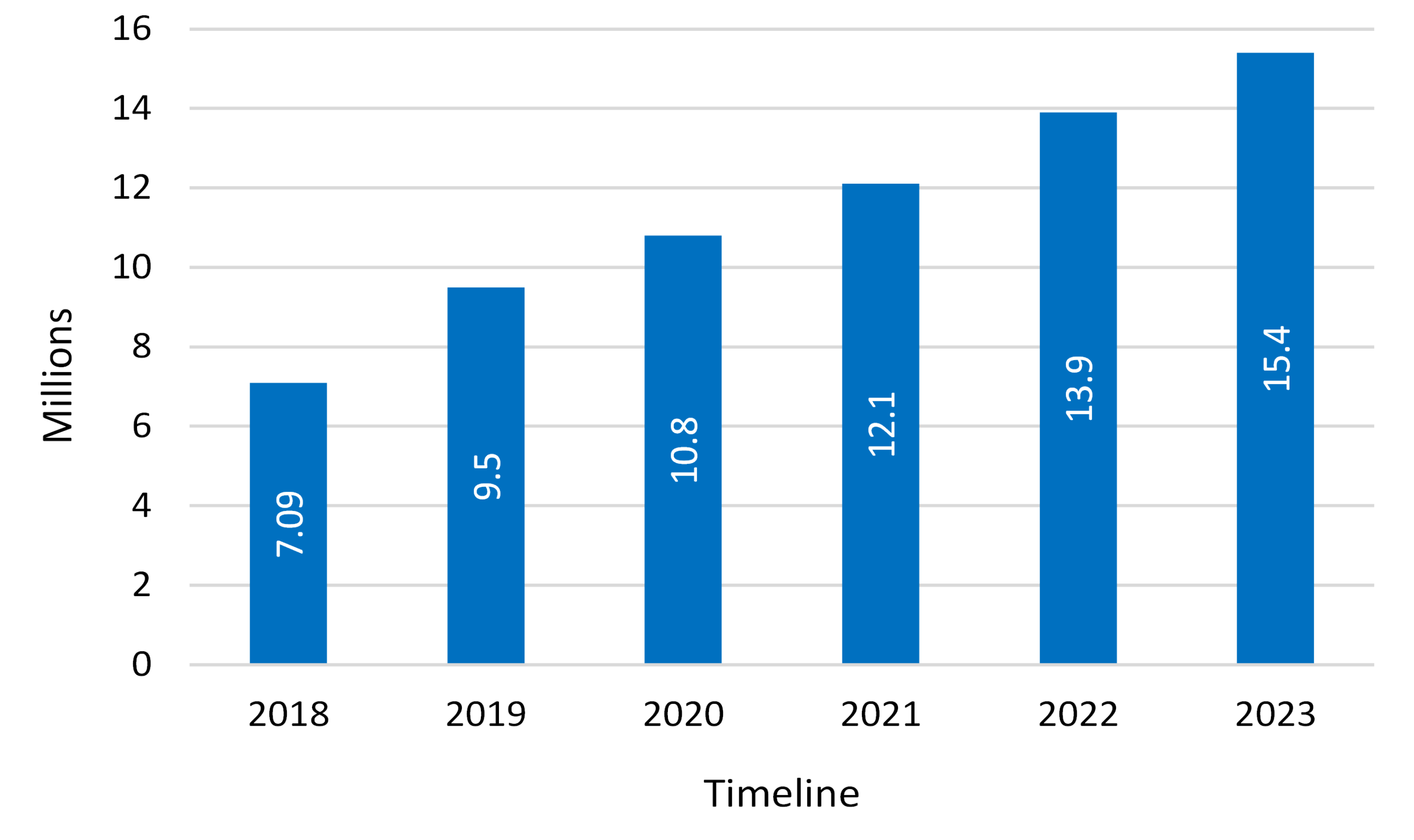}}
  \caption{Global Trend of DDoS Attacks 2018-2023 \cite{Cisco1}}
  \label{fig:cisco1}
\end{figure}

In the current era, security is the major concern of IoT \cite{hossain2019application}. Denial of service (DoS) attacks and distributed denial of service (DDoS) attacks have been reported as the most common attacks on IoT devices and network \cite{nguyen2019search}. A DoS attack is a malicious attempt done by an attacker using a single source to make a service or network resources inaccessible to legitimate users. When a DoS attack is launched using multiple distributed sources, it is called a DDoS attack.  The DoS and DDoS attacks are increasing rapidly both in frequency and intensity with an average of 28.7K attacks per day \cite{datto,jonker2017millions}. Recently, Neustar’s report of cyber threats and trends \cite{neustar} revealed that the DDoS attacks have been increased 200\% in frequency while 73\% increased in volume during the first six months of 2019 as compared to the same period in 2018 \cite{compari}. Fig. \ref{fig:cisco1} depicts the surging trend of DDoS attacks as anticipated in Cisco’s annual Internet report, 2018–2023 \cite{Cisco1}. It can be observed that by 2023, the total count of DDoS attacks would become double, i.e., 15.4 million as compared to 2018. Hence, there is a crucial need for developing such solutions which can effectively detect and devastate the DoS and DDoS attempts.  

So far, firewalls, intrusion detection systems (IDS) and intrusion prevention systems (IPS) are used as major security shields to protect the IoT devices and network from the cyber-attacks. However, the traditional firewalls, IDS and IPS cannot defend against the complex DDoS attacks \cite{nguyen2019search,khalaf2019comprehensive, ghazanfar2020iot, sharafaldin2019developing} as most of them filter the normal and suspicious traffic based upon the static predefined rules. However, the IDS and IPS that filter the intrusive attempts using artificial intelligence (AI) techniques are more reliable and effective as compared to the static predefined rules \cite{khalaf2019comprehensive}.
 
The traditional IDS use signatures or deep packet inspection (DPI) techniques for detecting malicious activities in the network. These techniques filter the packets based upon the packet content and header information. Unfortunately, such techniques have poor performance and become a bottleneck when deployed on high bandwidth and high-speed backbone links \cite{khalaf2019comprehensive}. Moreover, these techniques fail to check packet contents when the encrypted traffic flows over the network \cite{liu2019predicting}. Although many machine learning (ML) based solutions have been proposed for IoT attack detection, the prediction power of a well-tuned deep learning model especially convolutional neural network (CNN) is much better and effective as compared to the ML models \cite{liu2019predicting,xiao2019intrusion}. During the past few years, deep residual network (ResNet) drastically captivated the attention of researchers due to its tremendous performance \cite{liu2019predicting}. Thereupon, in this work, we used ResNet \cite{resnet_he2016deep} for IoT DoS and DDoS attack detection. 

No matter, the deep learning models especially CNN models have achieved high significance due to their efficient performance in image processing and computer vision field \cite{li2017intrusion}. However, these CNN models are also being used for detecting the network attacks.
Liu \textit{et al.} \cite{liu2019predicting}, proposed a CNN-based approach to detect the malicious traffic from NetFlow data. The authors first encoded the features then applied feature correlation and converted the data into images through surrounding correlation matrix. Finally, they fed these generated images to the deep learning models. Among these models, residual network (ResNet) \cite{resnet_he2016deep} outperformed the other models.
Likewise, Salman \textit{et al.} \cite{salman2019machine} devised a framework for IoT device identification and attack detection. The authors used a self-generated dataset of seven IoT devices and evaluated the processed framework using two machine learning and three deep learning networks out of which a machine learning model, i.e., Random Forest outperformed.
The authors in \cite{xiao2019intrusion} revealed that ResNet \cite{resnet_he2016deep} is prone to overfit in case of low dimensional and small size dataset due to which ResNet-based IDS do not perform well. To combat this challenge, the authors reconstructed the ResNet \cite{resnet_he2016deep} model by simplifying the residual block. The experiments proved that the simplified ResNet performed better as compared to the actual ResNet \cite{resnet_he2016deep} for low dimensional data.

The authors in \cite{liu2019novel} claimed that CNN best performs on images while the network traffic datasets are in non-image form. In order to efficiently use the potential CNNs for detecting the network intrusions, the authors proposed a methodology to convert the network traffic into a three-dimensional (3D) image. For this, the authors used a publicly available dataset, i.e., NSL-KDD dataset \cite{NSL-KDD}, applied fast Fourier transformation (FFT) onto it, converted it into 3D images and then passed it to a state-of-the-art CNN model to detect the network intrusions. Likewise, Li \textit{et al.} \cite{li2017intrusion} converted NSL-KDD dataset \cite{NSL-KDD} feature values into binary vectors using a binary encoding scheme then transformed these vectors into images. These images were fed into two deep neural networks. The authors concluded that CNN models show better performance as compared to the machine learning methods.
 
Although the potential of CNN models is being used for developing intrusion detection systems, these CNN models do not perform efficiently when trained on non-image dataset \cite{liu2019novel}. Hence, there is a need for developing such a mechanism that transforms the network traffic into a representable form on which CNN models perform efficiently. Usually, the network traffic datasets are in low dimensional form, i.e., either in .pcap format or in .csv or .txt format. While the CNN models are designed and widely known for solving image processing and computer vision problems. Therefore, the CNNs especially, ResNet \cite{resnet_he2016deep} do not perform well or overfit when trained on low dimensional and small-scale dataset \cite{xiao2019intrusion}. In order to handle this issue, we proposed a methodology to convert the network traffic captures, i.e., non-image data into a representable form, i.e., image form and trained a state-of-the-art CNN model over the converted data in order to better detect the DoS and DDoS attack patterns.

The existing IoT attack detection systems are unable to detect the latest DoS and DDoS attacks \cite{khalaf2019comprehensive, ghazanfar2020iot, sharafaldin2019developing}. The reason is that most of them are trained over either outdated datasets or the datasets used for training the proposed solutions do not include the modern reflective DDoS attacks like Network BIOS (NetBIOS) attack, Network Time Protocol (NTP) attack, Simple Service Discovery Protocol (SSDP) attack, UDP Lag (delay) attack, etc. In this work, we used the latest DDoS attack dataset, i.e., CICDDoS2019 \cite{sharafaldin2019developing} which contains 11 types of DoS and DDoS attacks and collected over a real-time network. Moreover, CICDDoS2019 \cite{sharafaldin2019developing} contains a large number of samples as compared to other network traffic datasets. In order to better detect the complex DoS and DDoS attacks, we used a state-of-the-art CNN model, i.e., ResNet \cite{resnet_he2016deep} which showed efficient performance in detecting the image patterns. For getting the efficient performance of ResNet \cite{resnet_he2016deep}, we proposed a methodology to convert the non-image network traffic dataset into three-channel images. 

A few works \cite{liu2019novel, liu2019predicting, li2017intrusion} converted the network traffic into images by applying some encoding method or some transformation technique like FFT to convert the data into image format. However, the proposed methodology is quite simple as we simply normalized the data instead of applying some computationally expensive encoding scheme or transformation technique like FFT. 
The proposed methodology showed better results for detecting the DoS and DDoS attack patterns as compared to the state-of-the-art work.

\begin{figure*}[t]
 \centering
  \centerline{\includegraphics[width=\linewidth]{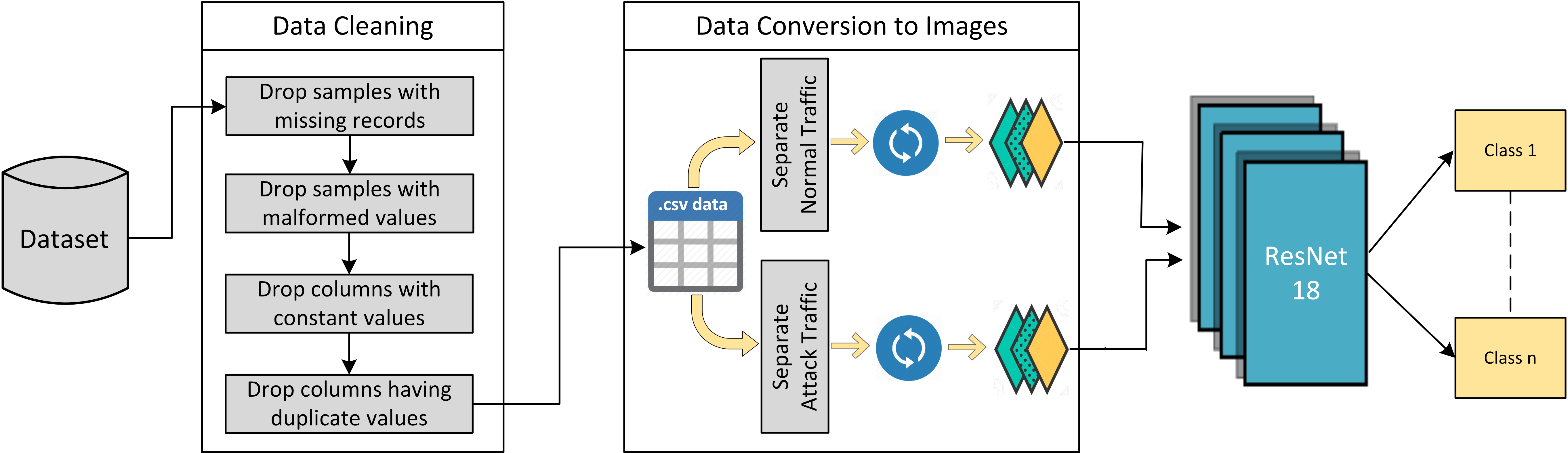}}
  \caption{Proposed Methodology For Detecting IoT DoS and DDoS Attacks using ResNet18}
  \label{fig:Proposed Methodology}
\end{figure*}

The rest of the paper is structured in the following manner: Section II defines the problem statement. Section III describes the proposed methodology for converting the non-image network traffic data into image form in order to better detect the DoS and DDoS attacks using a state-of-the-art CNN model. Section IV presents the results of the proposed methodology and shows how the proposed methodology outperforms state-of-the-art work. Lastly, Section V concludes the paper.

\section{Problem Statement}
DoS and DDoS attacks are the most common attacks in IoT \cite{nguyen2019search}. The existing solutions are unable to detect the complex DoS and DDoS attacks \cite{nguyen2019search,khalaf2019comprehensive, ghazanfar2020iot, sharafaldin2019developing}. The reason is that most of them are trained over either outdated datasets or the datasets which do not include the modern reflective DDoS attacks like NetBIOS, NTP, SSDP, UDP Lag, etc., \cite{sharafaldin2019developing}. The potential of deep learning models especially convolutional neural networks (CNNs) is being used for detecting the network attacks. However, CNNs do not perform efficiently when trained on non-image or low dimensional dataset \cite{liu2019novel, xiao2019intrusion} since CNNs are specially designed to solve computer vision problems. The network traffic datasets exist either in .pcap format or in .csv or .txt format. Therefore, to efficiently utilize the potential of CNNs for the DoS and DDoS attacks detection, the network traffic data should be converted into image form.

\section{Proposed Methodology}

The proposed methodology consists of four key steps which include: data acquisition, data cleaning, data conversion and attack pattern recognition. Fig. \ref{fig:Proposed Methodology} provides an overview of the proposed methodology. The first step of the proposed methodology is to acquire network traffic data. Once the data is acquired, it is preprocessed in order to get the refined data. During the preprocessing, we will perform two major steps, i.e., data cleaning and conversion of cleaned data into three-channel images. The final step is to train and test the CNN model over the preprocessed data in order to evaluate the performance for detecting the DoS and DDoS attack patterns. All these steps are explained in the following subsections. 

\subsection{Data Acquisition}

The data acquisition is the first step of the proposed methodology to acquire both normal and attack traffic. Generating extensive normal and attack traffic by setting up a real-time network is an onerous task as it requires significant network resources, diversity of network normal and attacks traffic captures, etc. Moreover, it is also a time and money consuming process to set up a huge network. However, one can get rid of this laborious job by using a publicly available network traffic dataset. In order to get a quality dataset, we analyzed some publicly available datasets based upon the following criteria:

\begin{itemize}
    \item The dataset must consist of real-time network traffic.
    \item The dataset must be extensive and versatile.
    \item The dataset must comprise of the latest DoS and DDoS attacks.
    \item The dataset should cover a variety of attack vectors.
\end{itemize}

We reviewed some publicly available datasets which include KDD-99 \cite{KDD}, NSL-KDD \cite{NSL-KDD}, DEFCON \cite{DEFCON}, CAIDA \cite{CAIDA}, UNSW-NB15 \cite{UNSW} and CICDDoS2019 \cite{sharafaldin2019developing}. Based on the above-mentioned criteria, for this work, we selected CICDDoS2019 \cite{sharafaldin2019developing} dataset. CICDDoS2019 \cite{sharafaldin2019developing} is the latest dataset which contains a large number of samples as compared to the other network traffic datasets. Moreover, it contains both inbound and outbound traffic of the latest DoS and DDoS attacks while most of the above-mentioned datasets are either outdated or contain limited attack scenarios of DoS attacks \cite{sharafaldin2019developing}. Furthermore, the CICDDoS2019 \cite{sharafaldin2019developing} dataset contains more than 80 network flow-related features and eleven types of latest DoS and DDoS attacks traffic collected over a real-time network. The details of these features are described in \cite{cicFeatures}.

\subsection{Data Preprocessing}
Once the data is acquired, the next stage is to preprocess the data in order to bring it in a refined form. During this stage, we performed three major steps, i.e., data cleaning, data conversion, and train test and validation split.

\subsubsection{Data Cleaning}
The acquired network traffic data was in .csv format which includes more than 80 flow features. In order to better train our model for attack pattern detection, we removed the unwanted features from the data set which are not useful for classifying the attack and normal traffic. These features include Flow ID, Source IP, Source Port, Destination IP, Destination Port, Protocol and Timestamp. As based upon these static features, one cannot decide whether a certain flowid, srcIP, etc., whenever found will always generate malicious or normal packets. That’s why we dropped such unwanted features and excluded them from our training set. 

Thereafter, we analyzed the whole dataset in order to deal with missing or malformed data. For this purpose, we first checked that which samples contain missing values, which samples have inadequate values like nan, -inf, +inf, etc. As we had a large number of samples in the dataset, so we dropped all those samples which comprise of missing or malformed values.

After that, we figured out the features which were either duplicate or entirely had a constant value in case of all labels. Such constant features are not useful for discriminating the attack or normal traffic and may decrease the performance of the machine learning model, if included in the training set. Therefore, we also dropped constant features from the training set. These features include Bwd PSH Flags, Fwd URG Flags, Bwd URG Flags, FIN Flag Count, PSH Flag Count, ECE Flag Count, Fwd Avg Bytes/Bulk, Fwd Avg Packets/Bulk, Fwd Avg Bulk Rate, Bwd Avg Bytes/Bulk, Bwd Avg Packets/Bulk and Bwd Avg Bulk Rate. On the other hand, the duplicate features are those which have similar values but has a different name. In the case of duplicate features, we keep the first original feature and dropped its duplicate feature. These features include RST Flag Count, Fwd Header Length, Subflow Fwd Packets, Subflow Fwd Bytes, Subflow Bwd Packets and Subflow Bwd Bytes. Finally, after the cleaning the data we left with 60 features which were unique and important.

\begin{figure}[t]
 \centering
  \centerline{\includegraphics[width=\linewidth]{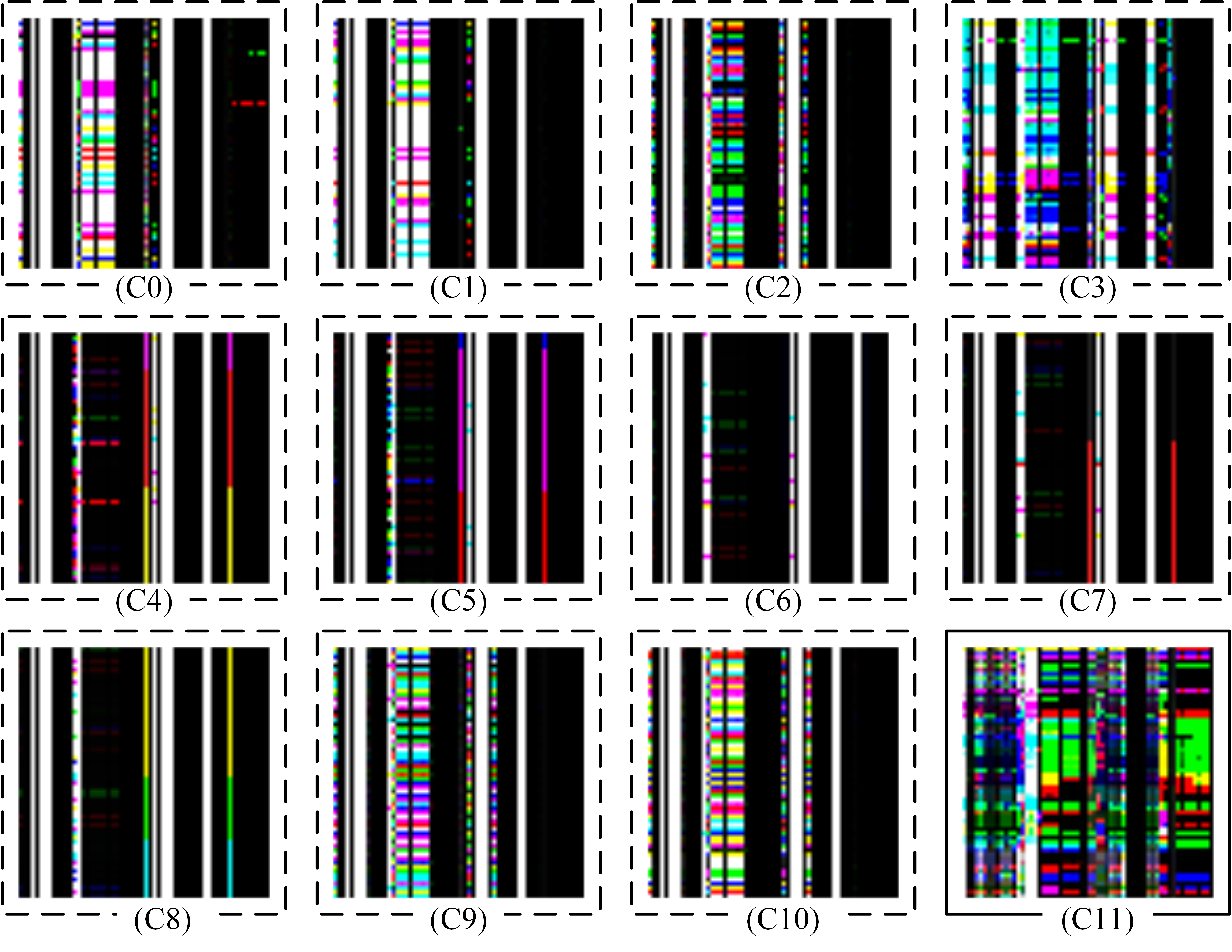}}
  \caption{Images obtained during Data Conversion Stage. (C0): Syn attack, (C1): TFTP attack, (C2): UDP Lag attack, (C3): DNS attack, (C4): LDAP attack, (C5): MSSQL attack, (C6): NetBIOS attack, (C7): NTP attack, (C8): SNMP attack, (C9): SSDP attack, (C10): UDP attack, (C11): Normal traffic}
  
  \label{fig:attack_images}
\end{figure}

\subsubsection{Data Conversion}

As mentioned earlier that CNN performs well when trained on an image-based dataset. Since the network traffic dataset is captured in non-image format, i.e., it can be either in .csv file, .txt file, or .pcap file. So, in order to get efficient results for network attack detection, we need to convert the network traffic data into image form. For this purpose, we normalized the dataset w.r.t each feature using (\ref{eq:normalization}).

\begin{equation}
\label{eq:normalization}
\textit{X'} = \frac{X - Min(X)}{Max(X)-Min(X)} \times 255
\end{equation}

In order to convert the network traffic into image form, we first separated the all normal and attack samples into two data frames as shown in Fig. \ref{fig:Proposed Methodology}. After that, we selected the chunk of 180 samples iteratively, converted them into an image of shape 60x60x3 in such a way that first 60 samples of each chunk were converted into image matrix of channel 1, next 60 samples of each chunk were converted into image matrix of channel 2 and the last 60 samples of each chunk were converted into image matrix of channel 3 and finally, mapped these matrices into RGB channels of an image. We converted these normal and attack samples matrices to images using the OpenCV library and labelled them accordingly. The same procedure was followed for each .csv file of the dataset until all the samples were converted into images. Fig. \ref{fig:attack_images} presents one sample of the converted images from each class, i.e., C0 to C11 which includes the 11 types of DoS and DDoS attacks along with normal traffic. In Fig. \ref{fig:attack_images}, the images enclosed inside a dot line box exhibits the attack images while the image inside the solid line box represents the normal image.

\subsubsection{Train and Test Split}
After converting the dataset into image form, the next step is to divide the dataset into a training set, validation set and testing set. In this regard, we randomly selected 2500 images from each category for testing and the images left were used for training the CNN model. Some of the classes had less than 2500 images, so we included all of them in the testing set.

\begin{figure}[t]
 \centering
  \centerline{\includegraphics[width=7cm, height=7cm]{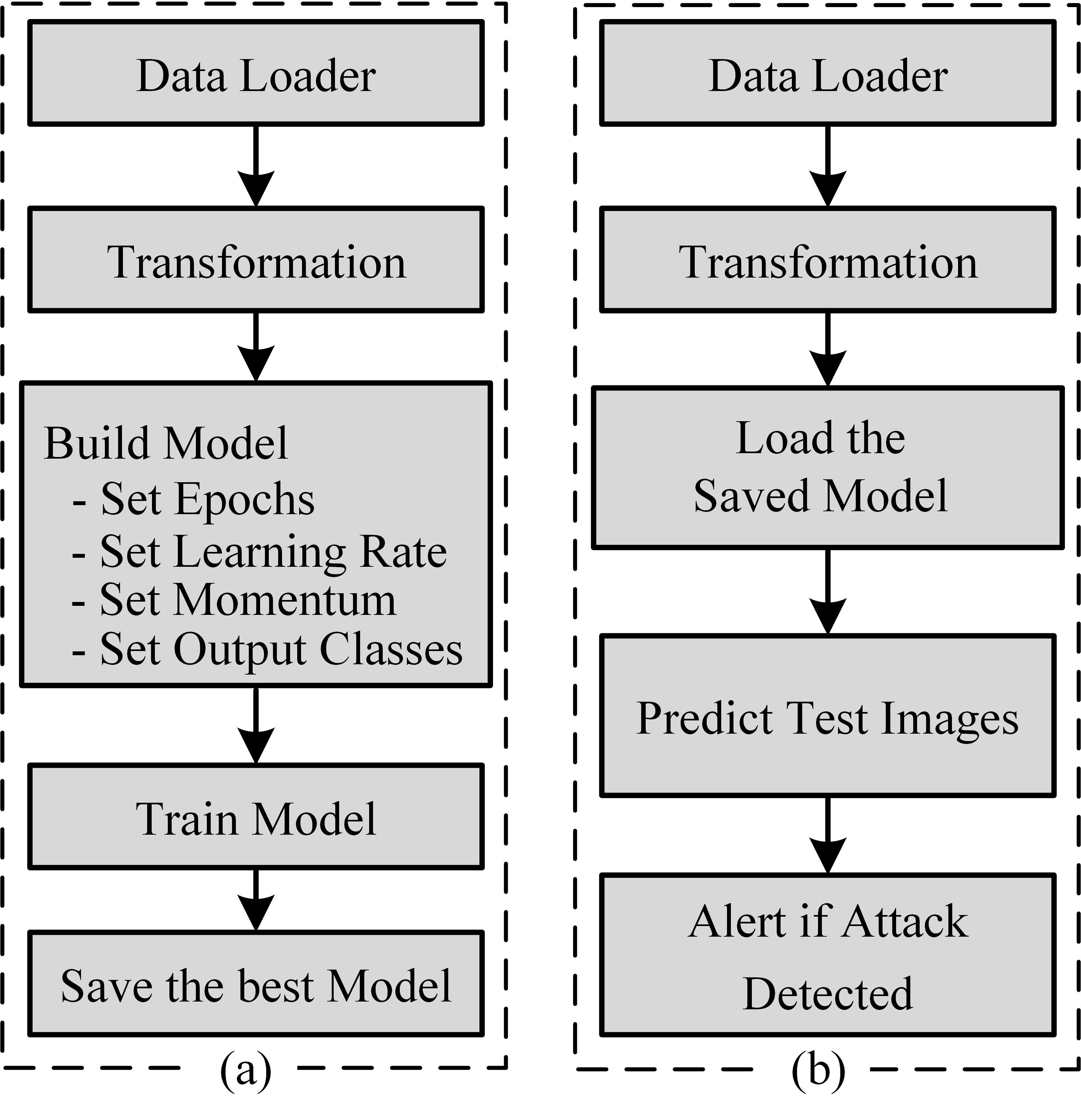}}
  \caption{Steps for (a) Training and (b) Testing ResNet18 Model}
  \label{fig:ResNet_Training}
\end{figure}

\subsection{Attack Detection}
Once the data is organized into train, test and validation set, the next step is to pass this data to the CNN model so that the model train itself on the given data, learn the attack and normal traffic patterns and validate itself. We used ResNet18 \cite{resnet_he2016deep} model which consists of 18 layers out of which 10 are convolution layers and 8 pooling layers. 

Fig. \ref{fig:ResNet_Training} – (a) shows the steps done for training the ResNet18 \cite{resnet_he2016deep} model on the network traffic dataset. We first loaded the train, validation set. The ResNet \cite{resnet_he2016deep} is designed to accept the images with size 224 x 224 \cite{liu2019predicting} while the preprocessed images had dimension 60 x 60 x 3. So, we transformed the preprocessed images into 224 x 224 x 3. After transforming the images, the next step is to build the model.

For building the ResNet \cite{resnet_he2016deep} model, we have to set some parameters as per use case. The original ResNet \cite{resnet_he2016deep} model had 1000 output classes but, in our use case, we set output class as 1 for binary classification while for multi-class classification, we set output classes as 12. So, we changed the last layer of ResNet \cite{resnet_he2016deep} model, to predict the given image according to our use case. Similarly, we also need to set some other parameters while building a model which includes epochs, learning rate, momentum and optimizer. We set the Resnet18 \cite{resnet_he2016deep} model with leaning rate 0.0001 with the momentum of 0.9, iterate over 10 epochs for binary classification and 50 epochs for multi-class classification with Stochastic gradient descent (SGD) optimizer.  
After building the models, we started the training of ResNet \cite{resnet_he2016deep} model over the prescribed parameters. While training the model, after each epoch we evaluated the model performance and saved the model weights if it gives the best accuracy. This evaluation process continued until the last epoch. Finally, we come up with the two best models, one for binary classification and other for multi-class classification. 

\begin{figure}[t]
 \centering
  \centerline{\includegraphics[width=\linewidth]{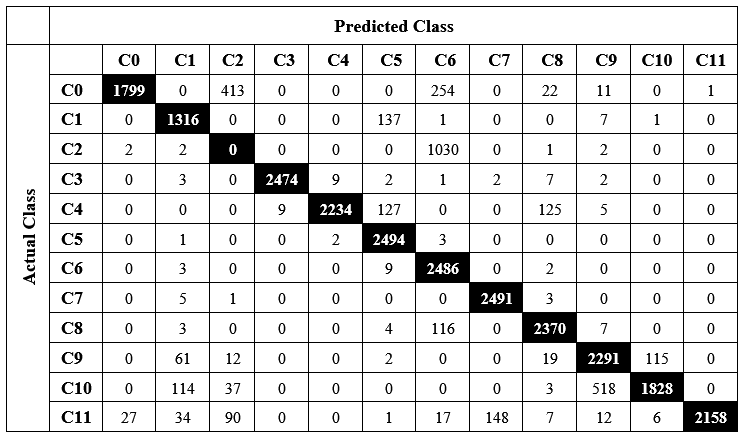}}
  \caption{Confusion Matrix obtained using Trained ResNet Model for DDoS Attack Detection}
  \label{fig:Conf_Matrix}
\end{figure}

\begin{table}[b]
\renewcommand{\arraystretch}{2}
\centering
\caption{Results Comparison}\label{results}
\begin{tabular}{|p{2.5cm}|p{1.5cm}|p{1.5cm}|p{1.6cm}|p{1.6cm}|}
\hline
\textbf{Method}  & \textbf{Precision} & \textbf{Recall} & \textbf{F1-Measure}               \\  \hline
Sharafaldin \textit{et al.} \cite{sharafaldin2019developing}  & 0.78  & 0.65   & 0.69                                           \\ \hline
Proposed  & 0.87  & 0.86   & 0.86                                         \\ \hline

\end{tabular}
\end{table}

Once the trained model with the best accuracy is saved, thenceforth, we need to test the trained models over the test set which remained unrevealed during the training phase. Fig. \ref{fig:ResNet_Training} – (b) illustrates the steps performed for testing the trained model. For testing the saved model, we first loaded the test images then transformed them into the dimension of 224 x 224 x 3 in a similar manner that we followed in the training phase. Afterwards, we loaded the saved models and passed the transformed images to it, so that it predicts whether the given images are normal or malicious in case of binary classification and if they are malicious then also predict their attack class in case of multi-class classification. Lastly, the predicted labels were compared with the actual labels in order to measure the performance of the trained model. The following section addresses the results achieved during the training and testing phase.

\section{Results and Discussion}
The performance of the proposed methodology is evaluated based on four commonly used performance metrics which include precision, recall, accuracy and F1-measure. In case of multi-class classification, all these parameters are calculated individually for each class from the confusion matrix which is shown in Fig. \ref{fig:Conf_Matrix}, then the average of each parameter is included in Table \ref{results}. These parameters are defined as:

\textbf{Precision} - {} It defines the ratio truly detected attacks and all packets that are classified as attacks. Mathematically, it is expressed in (\ref{eq:precision}): 

\begin{equation}
\label{eq:precision}
\textit{Precision} = \frac{TP}{TP+FP} \times 100
\end{equation}

\textbf{Recall} - {} It is the ability of the system to correctly detecting the attack upon the occurrence of the security breach. It is also called as the true positive rate. Mathematically, it is described in (\ref{eq:recall})

\begin{equation}
\label{eq:recall}
\textit{Recall} = \frac{TP}{TP+FN} \times 100
\end{equation}

\textbf{Accuracy} - {} It is defined as the ability of the system to correctly classify the attack packet as an “attack packet” and normal packet as a “normal packet”. It tells about the ratio of correct predictions with respect to all samples. Mathematically, it is expressed in (\ref{eq:accuracy}):

\begin{equation}
\label{eq:accuracy}
\textit{Accuracy} = \frac{TP+TN}{TP+FN+TN+FP} \times 100
\end{equation}

\textbf{F1-Score} - {} It is  defined as the harmonic mean of precision and recall. Mathematically, it is represented in (\ref{eq:f1-measure}): 

\begin{equation}
\label{eq:f1-measure}
\textit{F1-Score} = 2 \times \frac{Precision \times Recall}{Precision + Recall} 
\end{equation}

\begin{figure}[t]
 \centering
  \centerline{\includegraphics[width=\linewidth]{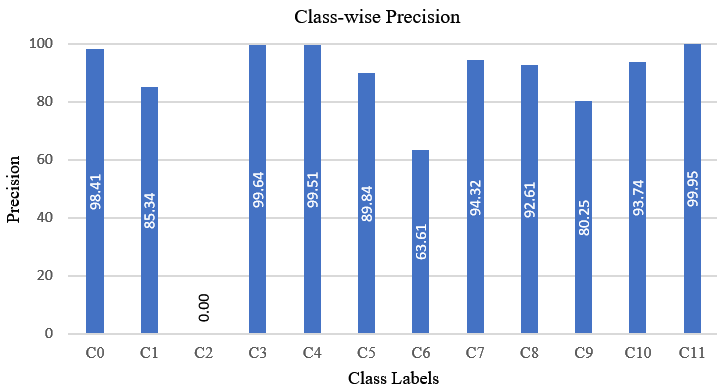}}
  \caption{Class-wise Precision obtained using Trained ResNet18 Model for DDoS Attack Detection}
  \label{fig:Class-wise_Precision}
\end{figure}

We evaluated the proposed methodology for DoS and DDoS attack detection based on the above-mentioned parameters during the training and testing phase for both detecting and recognizing the inbound and outbound DoS and DDoS attacks in IoT networks. In case of binary classification, the proposed methodology achieved 99.99\% accuracy for detecting the DoS and DDoS attacks. While in the case of multi-class classification, the proposed methodology achieved 87.06\% accuracy. Fig. \ref{fig:Class-wise_Precision} illustrates the percentage of correctly predicted attacks of 11 classes. It can be observed that Syn (C0), DNS (C3), LDAP (C4) attacks and normal traffic (C11) are detected with the highest precision, while UDP Lag attack is misdetected as NetBIOS attack. However, if we look at it in binary class perspective, then UDP Lag attack is correctly predicted as an attack.  

Overall, the proposed methodology using ResNet-18 \cite{resnet_he2016deep} model for multi-class classification, showed 88.56\% accuracy with a loss of 0.386 during the training phase. While upon testing, it showed the accuracy of 87.06\%. We also compared the results of our proposed methodology with a state-of-the-art solution in Table \ref{results}. It can be noticed that for detecting DoS and DDoS attack patterns, the proposed methodology exhibited 9\% more precision as compared to the state-of-the-art solution. Furthermore, it achieved 21\% higher recall, i.e., true positive rate and 17\% higher F1-score as compared to the state-of-the-art solution which was proposed on the same dataset.

\section{Conclusion}
The recent cyber-attack statistics reveal that denial of service (DoS) and distributed denial of service (DDoS) are the most occurring attacks in IoT networks and devices which are rising both in frequency and intensity by lapse of time. The convolutional neural network (CNN) models have gained a lot of significance in image classification tasks due to their outstanding performance. However, these models do not perform well when trained on a non-image dataset as they are designed to find the patterns from images. In order to use the potential of CNN models, in this work, we proposed a methodology to convert the non-image network traffic dataset into three-channel image form. Thereafter, we trained a state-of-the-art CNN model, i.e., ResNet over the transformed dataset and analyzed its performance for detecting the latest DoS and DDoS attacks. The proposed methodology is simple as compared to the existing works in such a way that it only normalizes the features and does not use any encoding scheme or transformation technique like fast Fourier transformation (FFT) to convert the preprocessed data into images. The proposed methodology achieved 99.99\% accuracy for detecting DoS and DDoS attacks. Furthermore, it attained 87\% precision for recognizing the 11 types of DoS and DDoS attacks which is 9\% higher as compared to the state-of-the-art.

\section{Declaration}
The authors declare no conflict of interest.

\bibliographystyle{IEEEtran}
\bibliography{bibliography.bib}

\end{document}